\documentclass{blois}

\def\be{\begin{equation}}
\def\ee{\end{equation}}
\def\bea{\begin{eqnarray}}
\def\eea{\end{eqnarray}}

\newcommand{\Photo}{}

\begin{document}
\vspace*{4cm}

\title{Top mass measurements in ATLAS and CMS}

\author{A. CASTRO$^\ast$}
\address{University of Bologna and INFN,\\
 Bologna, Italy\\
$^\ast$ On behalf of the ATLAS and CMS Collaborations}

\maketitle\abstract{Top quarks are produced copiously at the LHC, and a variety of related measurements has been made in the recent years by the two collaborations ATLAS and CMS. 
The most recent measurements of the top quark mass by the two collaborations are reported here. The top quark mass has been measured with a relative uncertainty smaller than 0.3\%, making the top quark the most accurately measured quark.}

\section{Introduction}
 At the LHC, the strong production of top quark-antiquark pairs is copious: more than 5 million $t\bar t$ pairs have been produced in $pp$ collisions at $\sqrt{s}=7$ and 
8 TeV (Run 1), yielding a $t\bar t\to W^+bW^-\bar b$ final state distinguished by the $W$'s decay into {\it dilepton}, {\it lepton+jets}, and {\it 
all-jets} channels. An additional million of top quarks has been produced singly.
\par
The top quark mass $M_t$ is an important free parameter of the Standard Model (SM) which can be measured directly from the observation of its decay products or, indirectly, comparing the $t\bar t$ cross section to theoretical expectations, yielding the so--called {\em pole mass}. 
Precise measurements of $M_t$, the $W$ mass $M_W$, and the Higgs boson mass $M_H$, are used to test the self-consistency of the SM~\cite{sm-global-fit}. In addition, 
top quarks might play a peculiar role in models for new physics~\cite{bsm-ref}.
Furthermore, $M_t$ and $M_H$ are related to the vacuum stability~\cite{vacuum-stability} of the SM, with the current value $M_H\approx 125$ GeV~\cite{pdg} corresponding to a near-criticality.
\par

\section{Measuring the top quark mass at the LHC}
The $t\bar t$ (or single top quark) events collected by ATLAS~\cite{atlas-ref} and CMS~\cite{cms-ref} have common physics 
signatures: high-$p_{\rm T}$ isolated 
leptons ($e$ or $\mu$); high-$p_{\rm T}$ jets, some of which associated to the hadronization of $b$ quark (i.e. $b$-jets); missing transverse momentum, $p_{\rm T}^{\rm miss}$,  associated to neutrinos. A recent addition to these physics objects regards the so-called {\em boosted jets}, which can be originated by top quarks produced at high $p_{\rm T}$.
All these physics objects are used to reconstruct the $pp\to t\bar t\to W^+bW^-\bar b$ final state (or the corresponding one for single top quark events), but ambiguities and permutations are to be considered for their mapping to the leptons/quarks of the final state. In addition, there is an uncertainty in the knowledge of the absolute value of jet energies, i.e. the so-called {\em jet energy scale} (JES), and the sharing of  $p_{\rm T}^{\rm miss}$ between multiple neutrinos.
\par
Given the number of top quarks produced, the statistical uncertainties are typically small, while the systematic ones are dominant. Thus,  it is important to study in detail the sources of uncertainty related to experimental effects, signal modeling and background modeling.
\section{Latest measurements by ATLAS and CMS}
We discuss now the most recent measurements of the top quark mass performed by the ATLAS and CMS Collaborations, with up to $20$ fb$^{-1}$ of integrated luminosity collected for $pp$ collisions at $\sqrt{s}=8$ TeV.  A very recent measurement, conducted by CMS with $2.2$ fb$^{-1}$ of $pp$ collisions at $\sqrt{s}=13$ TeV, is also reported.
\subsection{ATLAS measurements}
\begin{figure}[htb]
\begin{tabular}{ccc}
~~~ & 
\begin{minipage}{0.4\textwidth}
\includegraphics[width=6.8cm]{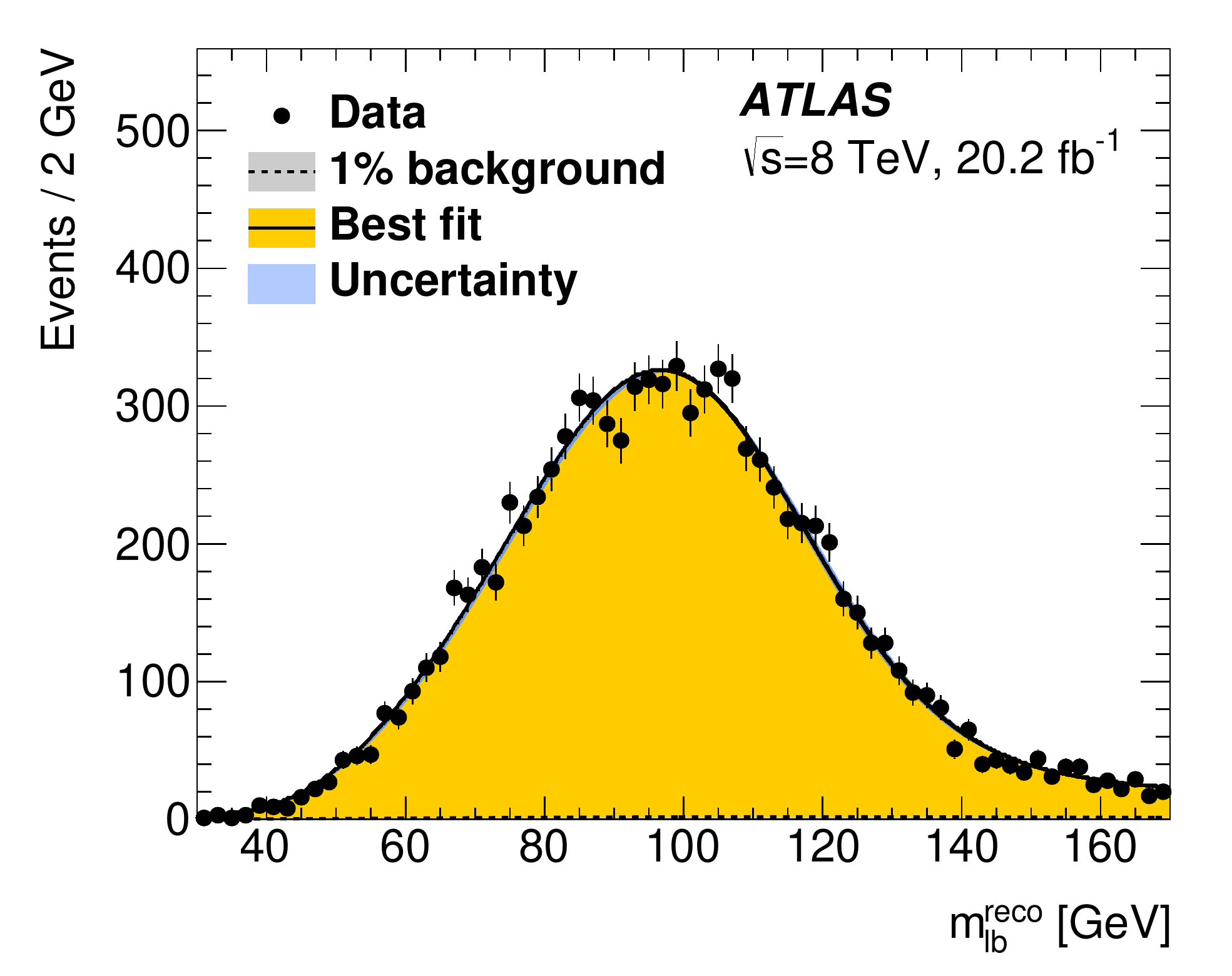}
\end{minipage}
& ~~~
\begin{minipage}{0.4\textwidth} 
\includegraphics[width=5.8cm]{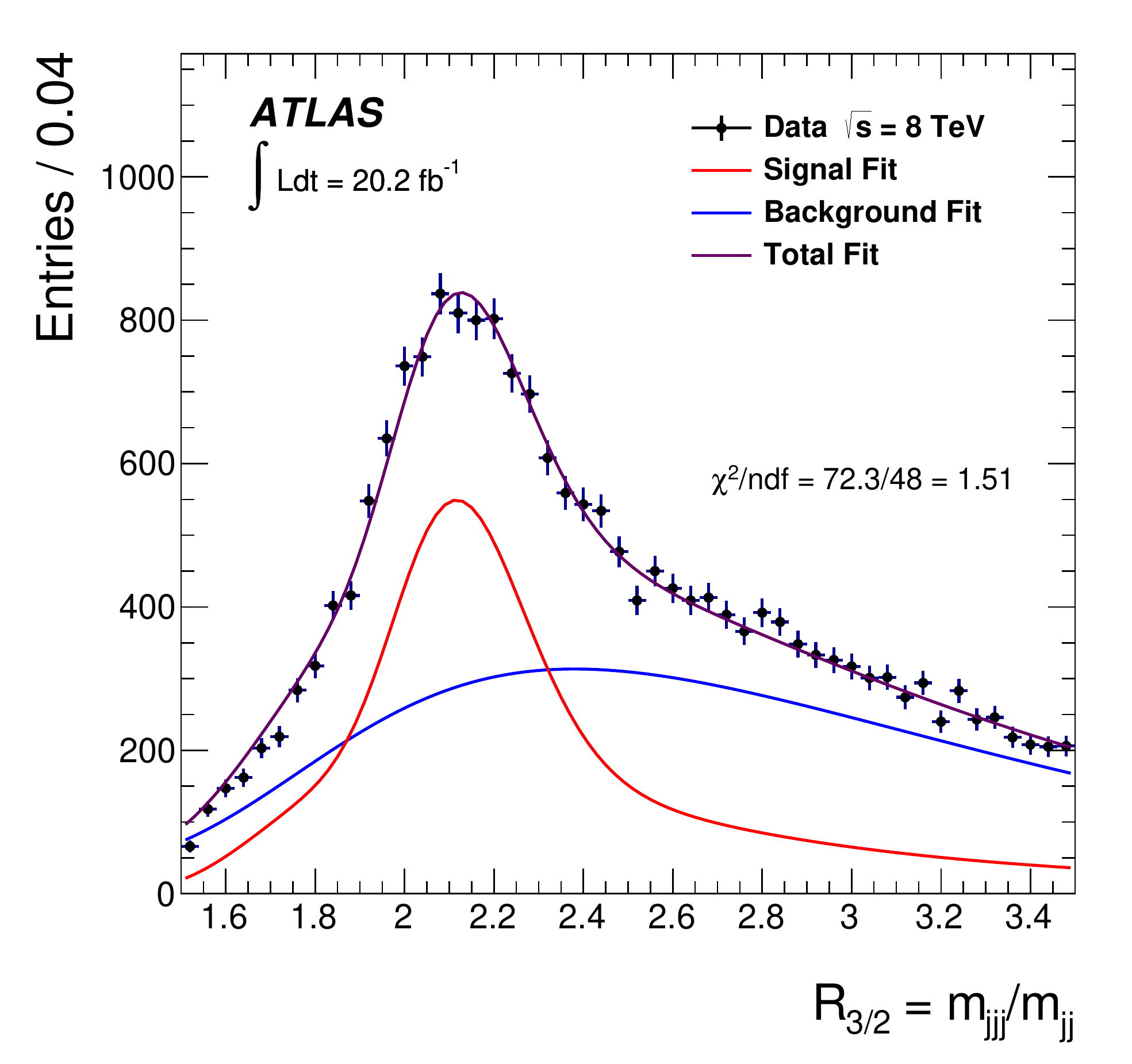}
\end{minipage}
\end{tabular}
\caption{ ATLAS template fits. Left: dilepton channel, $m_{\ell b}$ fit$^{~7}$. Right: all-jets events, $R_{3/2}$ fit$^{~8}$.}
\label{mass-atlas}
\end{figure}

{\bf Dilepton channel}, selecting events with two leptons ($e$ or $\mu$), at least two jets, one of which $b$-tagged.
For this channel ATLAS recurs to a  template method~\cite{atlas-dil-ref}. Because of the presence of two undetected neutrinos, the top quark mass cannot be reconstructed and what is used instead is the invariant mass $m_{\ell b}$ of the charged lepton and the $b$-tagged jet. The distribution of  $m_{\ell b}$  is used as template, exhibiting a dependence on $M_t$ parametrized with analytical functions. The  value of  $M_t$ returned by the template fit, see Fig.~\ref{mass-atlas} (left), amounts to $172.99\pm 0.41\,({\rm stat})\pm 0.74\,({\rm syst})$ GeV, with a total uncertainty of 0.85 GeV ($0.49\%$). The systematic uncertainty is dominated by contributions due to the  JES  uncertainty (0.54 GeV), $b$-jet corrections (0.30 GeV), and ISR/FSR effects (0.23  GeV).

\par\noindent
{\bf All-jets channel}, for events with at least 6 jets, including at least 2 $b$-tagged jets.
Also in this case ATLAS applies a template method~\cite{atlas-allj-ref} using as reference distribution the ratio $R_{3/2}$ between the invariant masses reconstructed from the jet triplets and doublets associated to top quark and $W$ decays.  The fit, shown in Fig.~\ref{mass-atlas} (right), returns an $M_t$ value of $173.72\pm 0.55\,({\rm stat})\pm 1.01\,({\rm syst})$ GeV, with a total uncertainty of 1.15 GeV ($0.66\%$). The systematic uncertainty is dominated by contributions from the modeling of the hadronization (0.64 GeV), the JES (0.60 GeV), and from the $b$-jet energy scale (0.34 GeV).

\par
\subsection{CMS measurements}
\begin{figure}[htb]
\begin{tabular}{ccc}
~~~~ & 
\begin{minipage}{0.4\textwidth}
\includegraphics[width=6.3cm]{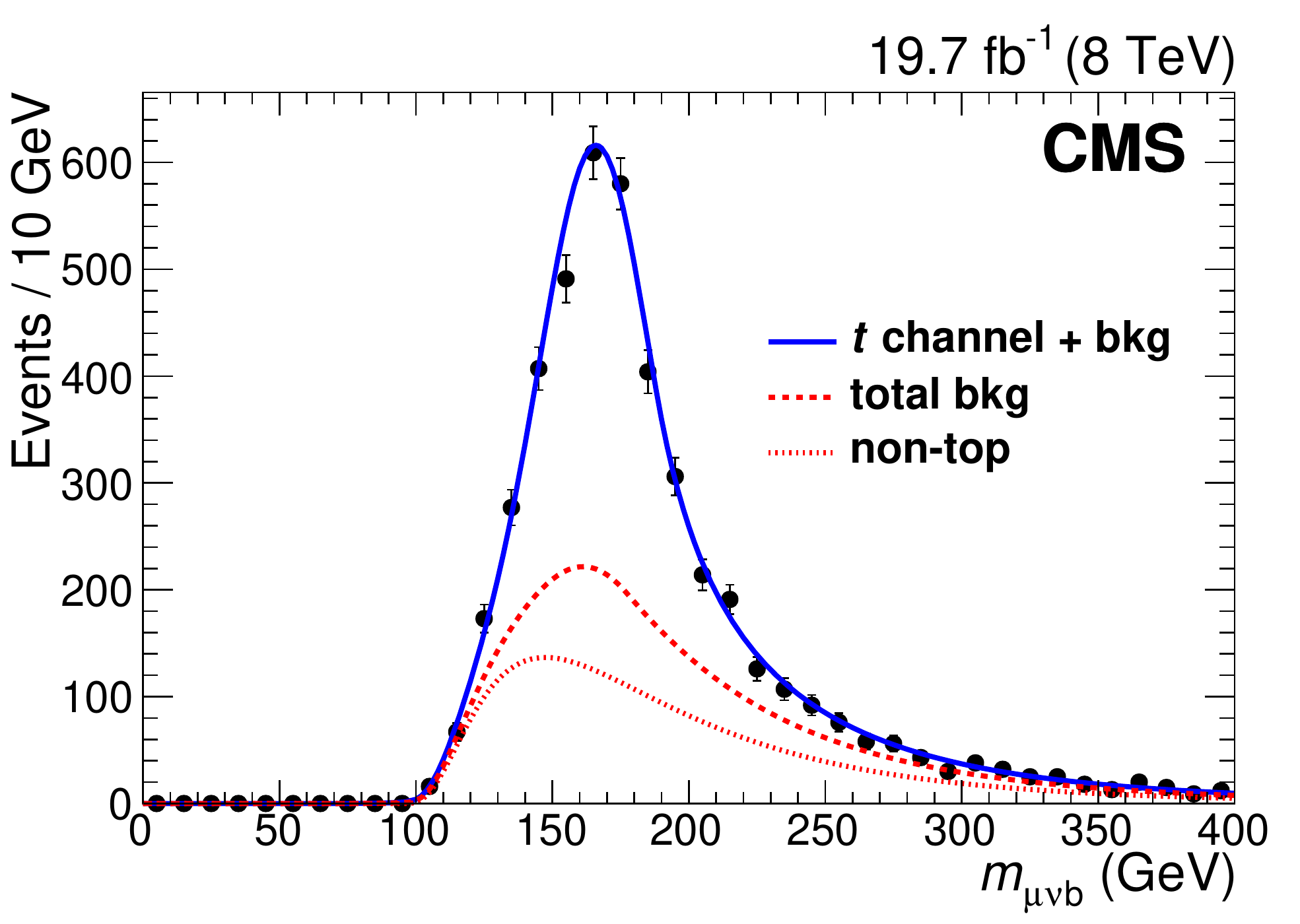}
\end{minipage}
& 
\begin{minipage}{0.4\textwidth} 
\includegraphics[width=6.3cm]{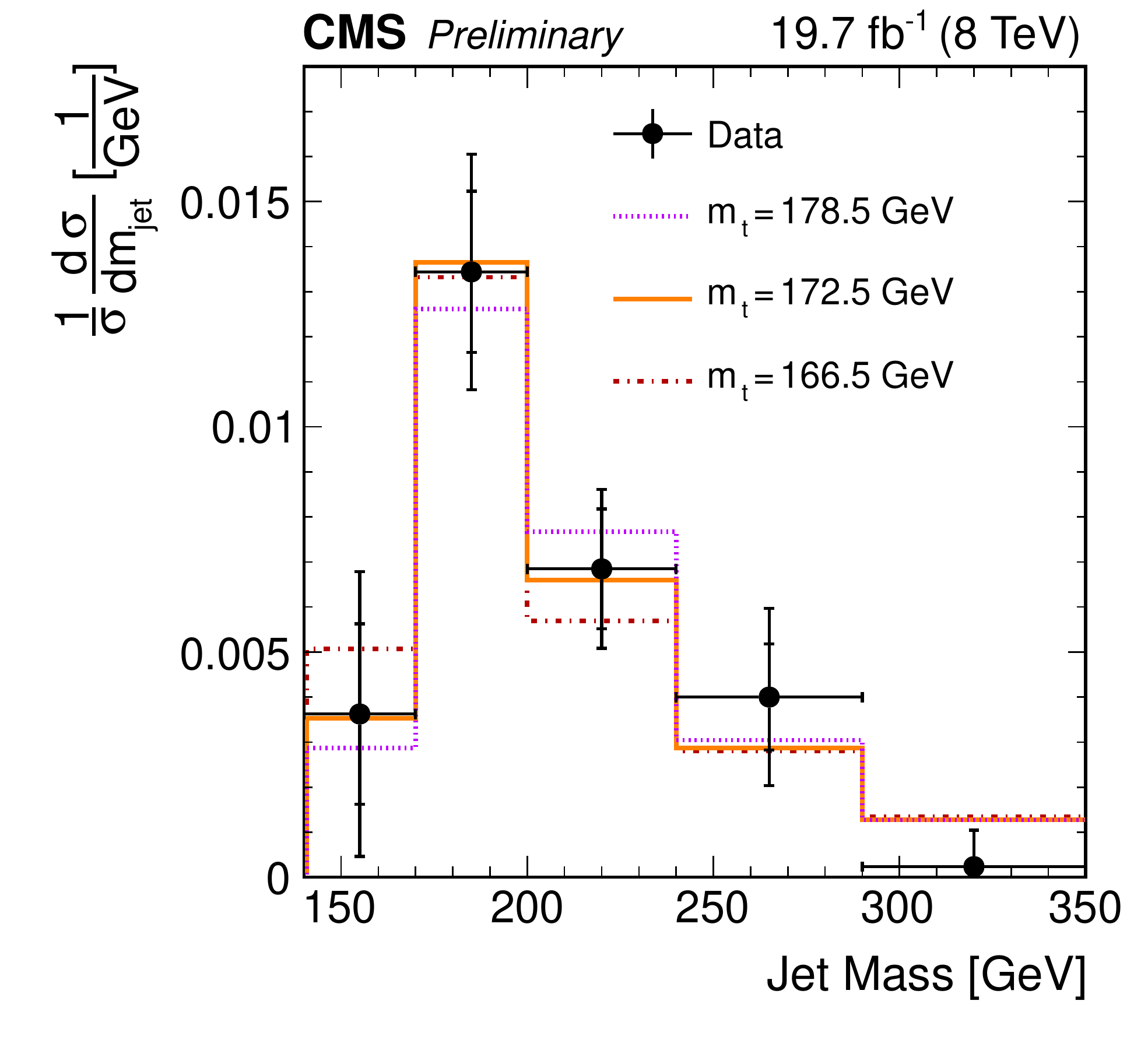}
\end{minipage}
\end{tabular}
\caption{ CMS events. Left: $m_{\mu\nu b}$ fit for single top quark, $\mu$ + jets events$^{~9}$. Right:  normalized $m_{\rm jet}$ differential cross section for boosted top quark events$^{~10}$. }
\label{mass-cms}
\end{figure}

{\bf Single top quark, $\mu$+jets}, using events with one muon, two jets, one of which $b$-tagged.
For this channel CMS recurs to a  template method~\cite{cms-single-ref}, using the invariant mass $m_{\mu\nu b}$ of the muon, 
 the $b$-tagged jet, and the neutrino whose momentum is inferred constraining the $\mu\nu$ invariant mass to $M_W$. 
The $m_{\mu\nu b}$  distribution  is used as template,
and described by analytical functions whose parameters are related to $M_t$. The $M_t$ value returned by the template fit, see Fig.~\ref{mass-cms} (left), 
is 
 $172.95\pm 0.77\,({\rm stat})^{+0.97}_{-0.93}\,({\rm syst})$ GeV, with a total uncertainty of 1.24 GeV ($0.72\%$).
 The systematic uncertainty is dominated by  the  JES uncertainty (0.68 GeV), the background modeling  (0.39 GeV), and the fit calibration (0.39 GeV). 

\par\noindent
{\bf Boosted top quark}, for events with one lepton ($e$ or $\mu$), at least 2 wide (boosted) jets, at least 3 narrow jets, including at least 1 $b$-tagged jet.
In this case CMS measures~\cite{cms-boost-ref} the differential cross section as a function of the boosted jet invariant mass  $m_{\rm jet}$, for jets with $p_{\rm T}>500$ GeV. The normalized differential cross section depends indeed on $M_t$, as shown in  Fig.~\ref{mass-cms} (right). A template fit returns a value $M_t=171.8\pm 5.4\,({\rm stat})\pm 3.0\,({\rm syst})\pm 5.5\,({\rm model})\pm 4.6\,({\rm theory}) $ GeV, with a total uncertainty of 9.5 GeV ($5.5\%$). This large systematic uncertainty is dominated by contributions from the signal modeling, but the method itself works. 
\begin{figure}[htb]
\begin{tabular}{ccc}
~~~ & 
\begin{minipage}{0.4\textwidth}
\includegraphics[width=6.3cm]{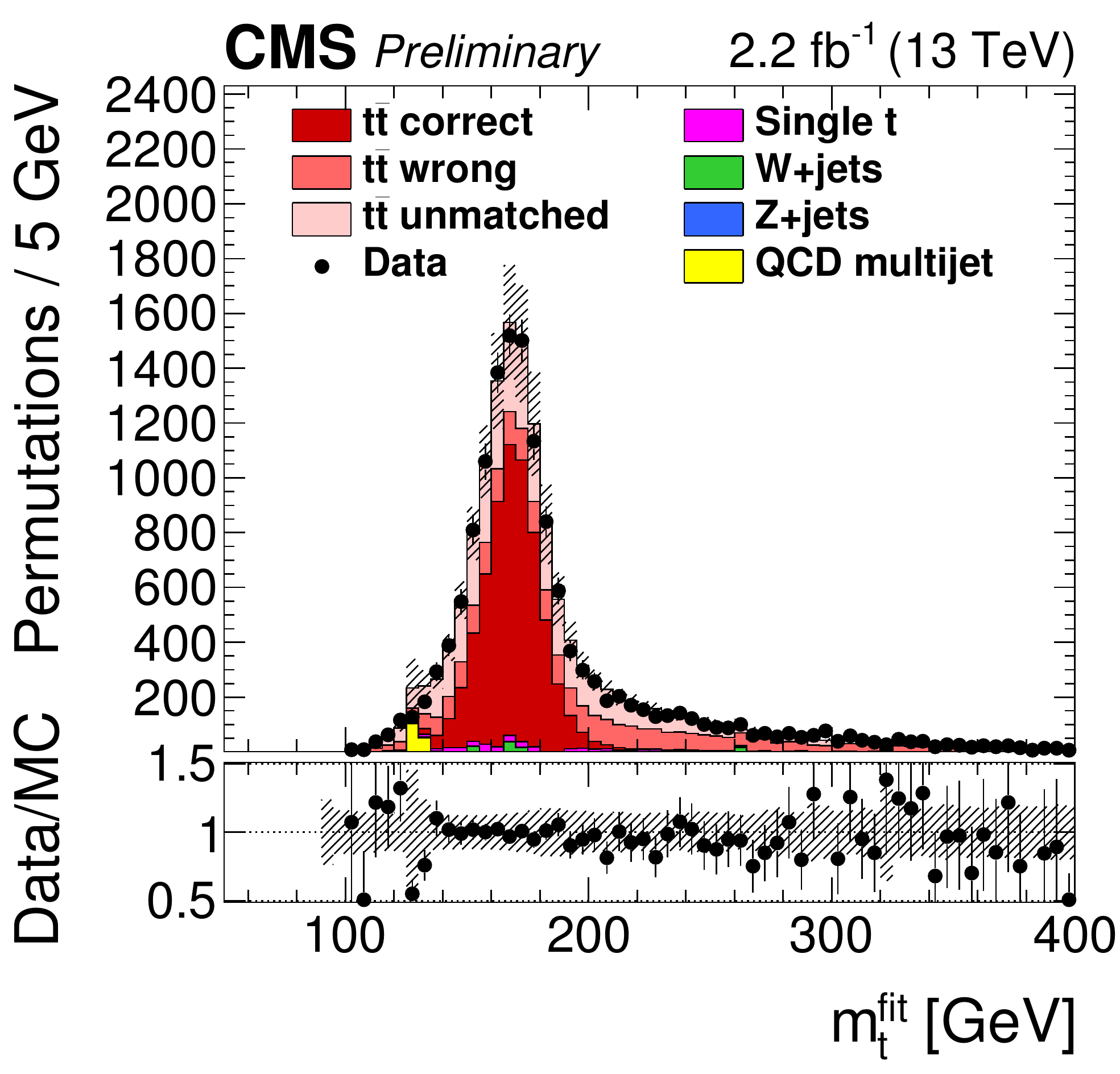}
\end{minipage}
& ~~~
\begin{minipage}{0.4\textwidth} 
\includegraphics[width=6.3cm]{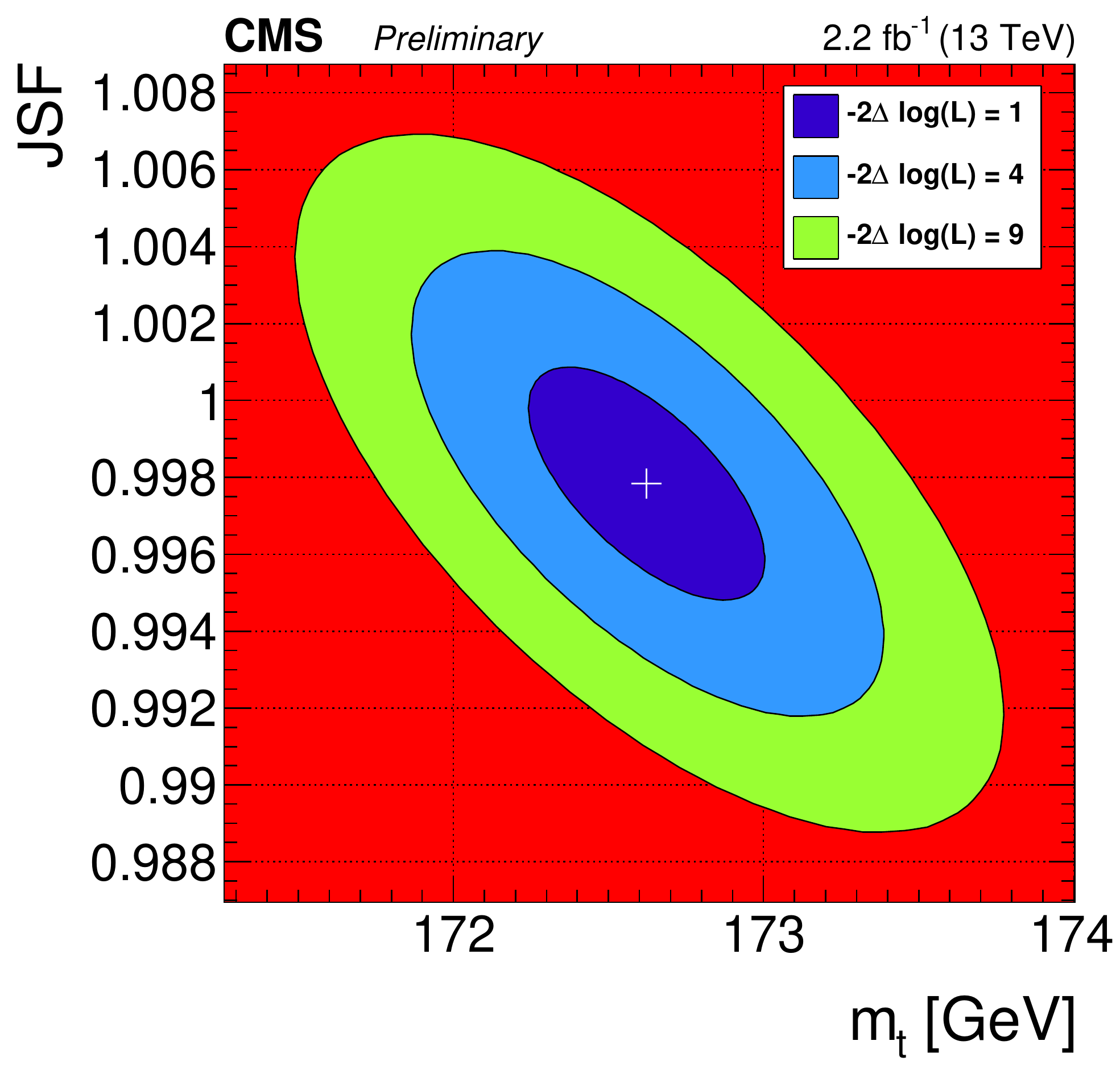}
\end{minipage}
\end{tabular}
\caption{CMS $\mu$+jets events at 13 TeV$^{~11}$. Left: distribution of  $m_t^{\rm fit}$. Right: 2-D contour plot $JSF$ vs $M_t$.}
\label{mass-cms-13}
\end{figure}

\par
{\bf $\mu$+jets channel at 13 TeV}, for events with one $\mu$,  at least 4 jets, including 2 $b$-tagged jets.
This very recent measurement~\cite{cms-13-ref} is based on  2.2 fb$^{-1}$ of $pp$ collisions at $\sqrt{s}=13$ TeV collected in 2015.
 The measurement is based on the so--called {\em ideogram method}.
 Starting from the kinematical reconstruction of the $WbWb$ final state, the method computes an event likelihood as a function of $M_t$ and of the reconstructed $W$ boson mass, convoluting Breit-Wigner (or similar) distributions with experimental resolutions. Multiple combinations for the jet-to-quark matching are considered with weights depending on the goodness of the fit. The signal purity is quite improved by a request on the fit probability, yielding a very peaked distribution for the reconstructed top quark mass $m_t^{\rm fit}$, as shown in Fig.~\ref{mass-cms-13} (left). The fit is based on 2-dimensional distributions of the $W$ boson and top quark reconstructed masses, and this allows  an in situ calibration of  a factor $JSF$ which modifies the default JES.
The values returned by the fit, see Fig.~\ref{mass-cms-13} (right), 
are $JSF=0.998\pm 0.010$ and  $M_t=172.62\pm 0.38\,({\rm stat}+JSF)\pm 0.70\,({\rm syst})$ GeV, with a total uncertainty of 0.8 GeV ($0.46\%$).
 The systematic uncertainty is  dominated by the flavor-dependence of the JES (0.41 GeV),  residual JES effects (0.30 GeV), and the parton-shower modeling (0.23 GeV).

\par
\subsection{Run 1 top quark mass combinations}
\begin{figure}[htb]
\begin{tabular}{ccc}
~~~ & 
\begin{minipage}{0.4\textwidth}
\includegraphics[width=7.5cm]{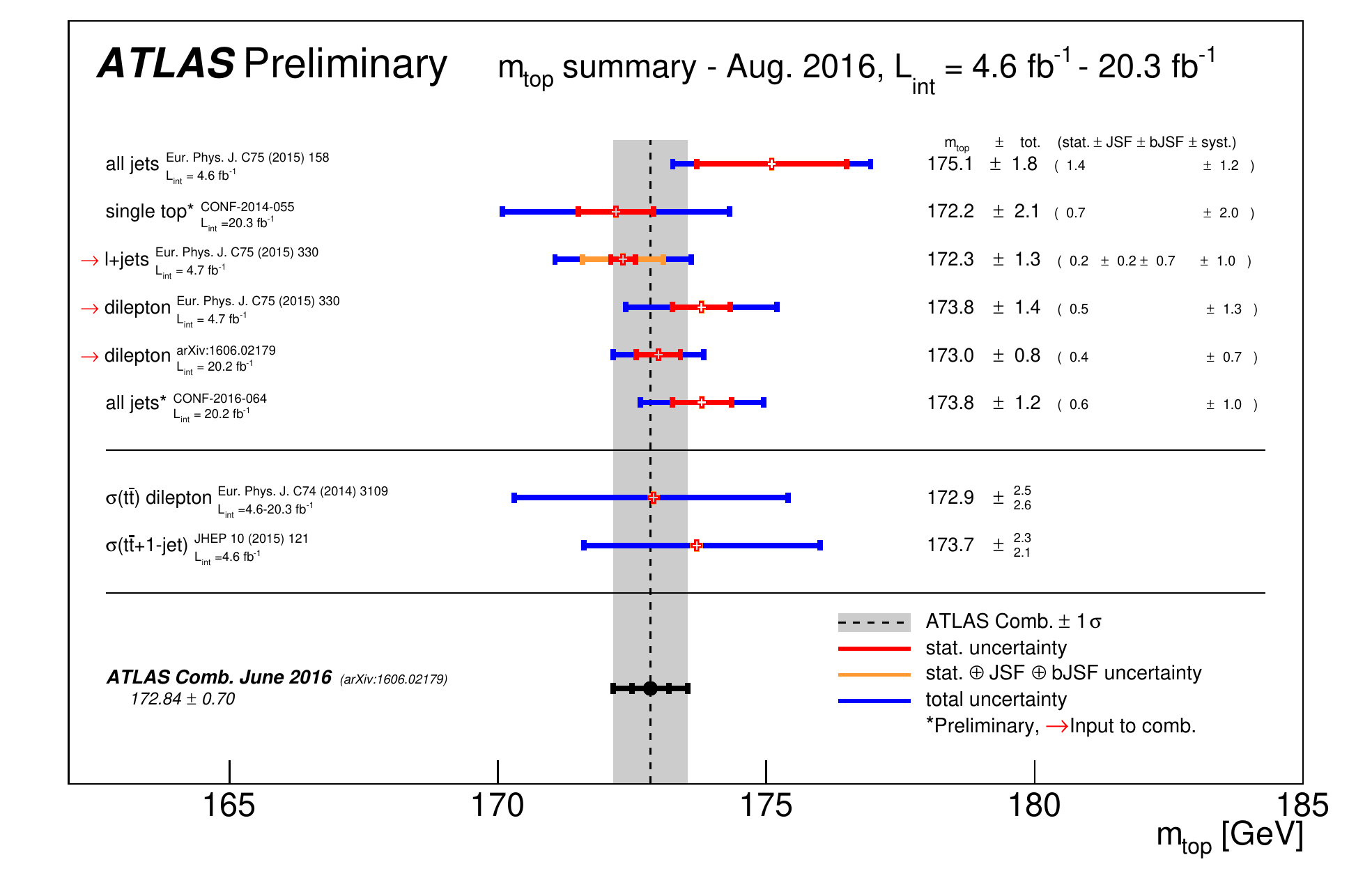}
\end{minipage}
  &  ~~~
\begin{minipage}{0.4\textwidth} 
\includegraphics[width=5.0cm]{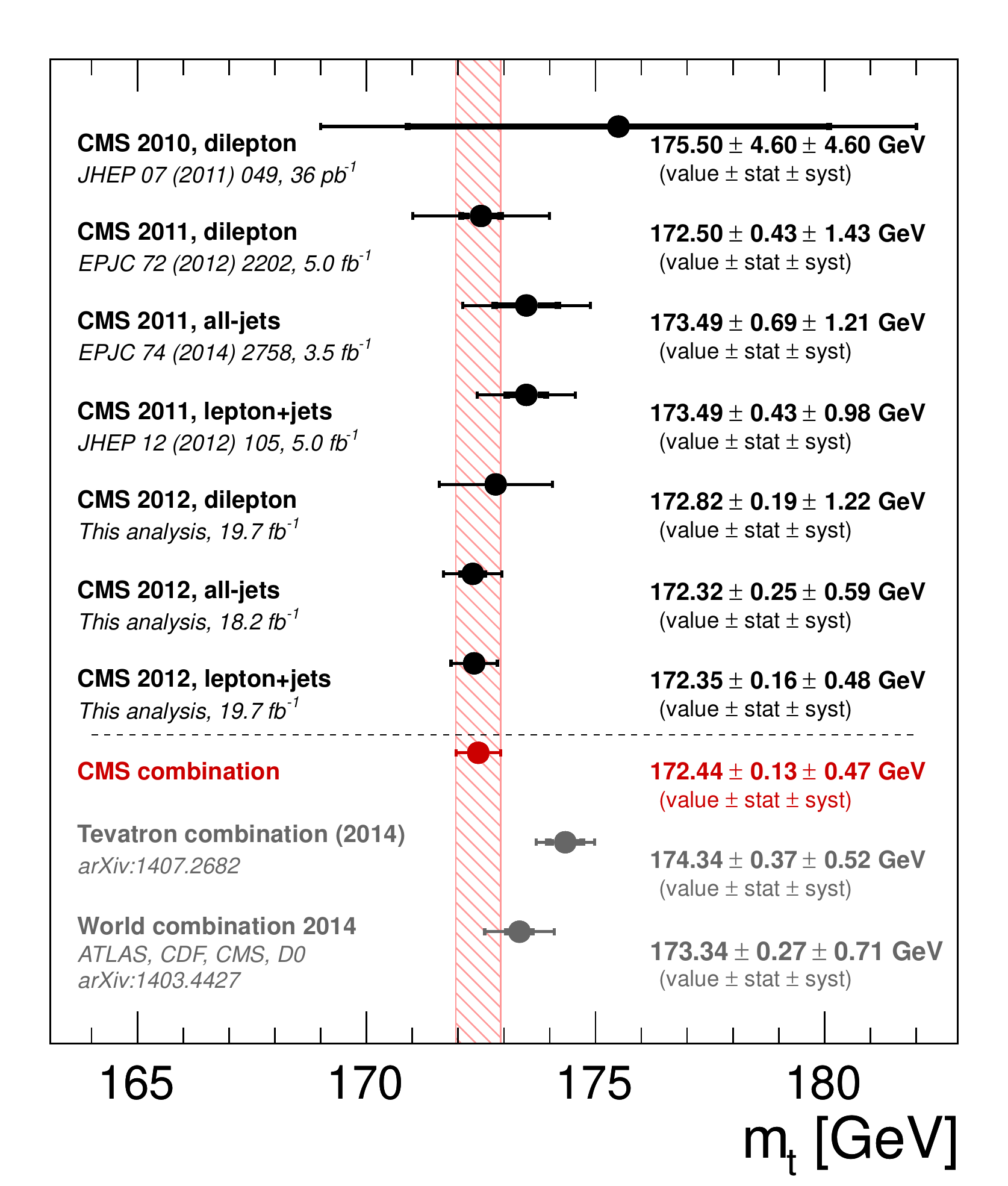}
\end{minipage}
\end{tabular}
\caption{Summary of $M_t$ measurements at 7 and 8 TeV. Left: ATLAS$^{~12}$. Right: CMS$^{~13}$.}
\label{mass-combos}
\end{figure}

{\bf ATLAS combination.}
ATLAS performed several mass measurements~\cite{atlas-ave-ref} at 7 and 8 TeV, as shown in Fig.~\ref{mass-combos} (left),  but the most accurate one comes from the combination of dilepton 
results at 8 TeV and dilepton and lepton+jets results at 7 TeV~\cite{atlas-dil-ref}.  The combination amounts to $M_t=172.84\pm 0.34\,({\rm stat})\pm 0.61\,({\rm syst})$ GeV, 
with a total uncertainty of 0.70 GeV (0.41\%). \par\noindent
 {\bf CMS combination.}
The mass measurements performed by CMS at 7 and 8 TeV, summarized in Fig.~\ref{mass-combos} (right), are combined~\cite{cms-ave-ref}  yielding a value  $M_t=172.44\pm 0.13\,({\rm stat})\pm 0.47\,({\rm syst})$ GeV, with a total uncertainty of 0.49 GeV corresponding to 0.28\%.
\par\noindent
 {\bf World combination.}
These combinations made individually by ATLAS and CMS are more precise than the 2014 world average~\cite{world-ave-ref} $M_t=173.34\pm 0.76$ GeV, which had a  total uncertainty of $0.44\%$. 
The inclusion of the newer LHC results will improve  the world average.

\section{Summary}
The top quark has been discovered 22 years ago, and since then the measurement of its mass has been pursued with a variety of channels and techniques.
The precision reached is impressive, smaller than $0.3\%$, thanks to the accumulation of data and refinements in the methodology. Improvements on the  precision are expected from ongoing and future measurements at the LHC. New measurements at increasing precision will help to shed light on fundamental  cosmological issues and on physics beyond the SM.
To meet these challenges it will be important to reduce the systematic uncertainties, mainly those related to signal modeling, through a better tuning of the parameters in the Monte Carlo generators improving the agreement with the data.

\end{document}